\documentstyle[12pt]{article}
\begin{document}             
\global\arraycolsep=2pt 
\baselineskip=0.82cm
%
\thispagestyle{empty} 
\begin{titlepage}    
\begin{flushright}UT-Komaba 96-29  \end{flushright}
                 
%
        
%
%
\begin{center}
{\Large\bf Spontaneous Symmetry Breaking in Discretized
               Light-Cone Quantization }
\end{center}                                   
            
\vspace{0.5cm}
              
\begin{center}
Kazunori ITAKURA $^{(a)}$ \footnote{E-mail: itakura@hep1.c.u-tokyo.ac.jp}  
 and
Shinji MAEDAN $^{(b)}$ \footnote{Permanent address: Department of
                      Physics,
                        Tokyo National College of Technology,
                  1220-2 Kunugida-machi, Hachioji-shi, Tokyo 193, Japan.
                   E-mail: maedan@tokyo-ct.ac.jp}             \\
\vspace{0.8cm}
{\sl $^{(a)}$ Institute of Physics,
          University of Tokyo,
            Komaba, Meguro-ku, Tokyo 153, Japan }
{\sl $^{(b)}$ Department of Physics,
          Tokyo Metropolitan University,
            Minami-Ohsawa, Hachioji-shi, Tokyo 192-03, Japan }
\end{center}                                               
            
\vspace{0.5cm}
              
\begin{abstract}
\baselineskip=0.82cm
\noindent
Spontaneous symmetry breaking of the light-front Gross-Neveu model is studied
   in the framework of the discretized light-cone quantization.
Introducing a scalar auxiliary field and adding its kinetic term, 
   we obtain a constraint on the longitudinal zero mode of the scalar field.
This zero-mode constraint is solved by using the $1/N$ expansion.
In the leading order, we find a nontrivial solution which gives the fermion 
   nonzero mass and thus breaks the discrete symmetry of the model.
It is essential for obtaining the nontrivial solution to treat adequately 
   an infrared divergence which appears in the continuum limit.
We also discuss the constituent picture of the model.
The Fock vacuum is trivial and an eigenstate of the light-cone
   Hamiltonian.
In the large $ N $ limit, the Hamiltonian consists of the kinetic term
   of the fermion with dressed mass and the interaction term of these fermions.
\end{abstract} 
\vfill            
\end{titlepage}
%
%
%
%
\setcounter{page}{1}
\section{Introduction}

One of the most important problems in QCD is to calculate the mass spectrum
   of hadrons, but this is a very difficult task due to, for example,
   the nonperturbative properties of the QCD vacuum such as quark 
   confinement and chiral symmetry breaking.
On the other hand, we know that the constituent quark model is a good 
   phenomenological model and that it succeeds to some extent 
   in obtaining the hadron spectrum.
The constituent quark model provides us with a very simple and 
   intuitive picture of hadron structure that  they are composed of 
   a few constituent quarks with relatively large masses
   ($\sim O( 10^2 \rm{MeV} ) $ for $u$, $d$ and $s$ quarks) 
   in a confining potential with a trivial vacuum.

In the light-cone (LC) quantized field theory \cite{rf:Dir,rf:rev}, 
   the vacuum state is trivial
   (if we simply ignore the longitudinal zero modes), and 
   the lower excitation can be approximated by the Fock states with small 
   particle number.  
 For example, if we apply the LC formalism to QCD, a mesonic state will be
   approximated as $|{\rm meson} \rangle \sim | q \bar{q} \rangle
   +~ | q \bar{q}~ , q \bar{q} \rangle $.
Indeed, in the 1+1 dimensional QCD on the LC, 
   the ground state of a meson is represented by a single quark-antiquark
   pair quite well \cite{rf:HorBroPau}. \ 
Therefore, the LC quantization is a promising method 
   to bridge a gap between QCD and the constituent quark
   model \cite{rf:WilRob}. \ 

Although the LC quantization has such attractive features, there arises 
   a question concerning the vacuum of QCD.
In the equal-time quantization formalism, the nonperturbative properties 
   such as the spontaneous chiral symmetry breaking and the quark confinement
   are attributed to the vacuum physics. 
The vacuum in the equal-time formalism is very complicated and difficult
   to determine.
How can we explain such complexity of the QCD vacuum in the light-cone
   quantized QCD ?~
In order to give answers to this question, many studies have been done using
   simple models such as the
   $ \phi^4 $ theory in $ 1+1 $ dimensions
   \cite{rf:HeiKruSimWer,rf:Rob,rf:PinSanHil}, 
   $ {\rm O}(N) ~ \phi^4 $ model \cite{rf:BorGraWer}
   or the Gross-Neveu
model \cite{rf:Hor,rf:ThiOht,rf:Pes,rf:KojSakSak,rf:Mae}. \ 
These models are known to exhibit spontaneous symmetry breaking 
   in the equal-time formulation.

 From these studies, it has been revealed that the longitudinal zero mode 
   (simply referred to as the 'zero mode') of the field operator
   plays an important role in realizing the spontaneous symmetry breaking with
   the trivial vacuum state.
In the discretized light-cone quantization (DLCQ) \cite{rf:PauBro}, 
   the momentum is discretized so that
   the zero mode can be treated clearly.
Because the scalar zero mode satisfies the so-called 
   ``zero-mode constraint," it is
   not an independent variable and is represented by other
   oscillation modes \cite{rf:MasYam}. \ 
If the zero-mode constraint possesses a solution 
   whose vacuum expectation value does not vanish, 
   such a solution leads us to the broken phase of the symmetry.
In addition to the constrained zero mode, there is another type of 
   zero mode which lives in the gauge field and is called the 
   ``dynamical zero mode" \cite{rf:KalPauPin}. \ 
This zero mode,
   that is an independent physical variable, is also expected to give rise to
   rich structure of the vacuum.

In this paper, we consider the light-front Gross-Neveu model in order
   to understand how the spontaneous symmetry breaking occurs in 
   DLCQ and to study the constituent picture of the model.
The reason why we choose the Gross-Neveu model \cite{rf:GroNev} is that
   it has many common properties with QCD, such as 
   spontaneous chiral symmetry breaking, asymptotic freedom,
   and the $ 1/N $ expansion.
Furthermore, in the Gross-Neveu model, fermions dynamically obtain a nonzero 
   mass $\sigma\neq 0$ and form a bound state $ \bar{\psi} \psi $ with mass
   $ 2 \sigma $  in the large $ N $ limit.
We particularly pay attention to problems of how the spontaneous chiral
   symmetry breaking is realized and how the zero mode affects the
   Hamiltonian in DLCQ.

The light-front Gross-Neveu model has been considered in
   several works in various
   approaches \cite{rf:Hor,rf:ThiOht,rf:Pes,rf:KojSakSak,rf:Mae}. \ 
In a previous work \cite{rf:Mae}  by one of the authors,
   an auxiliary scalar field was introduced and 
   the fermion (we shall call it quark) fields were integrated   
   over.
The zero mode of the auxiliary field was found to play a
    significant role for the spontaneous symmetry breaking.
Although this approach makes the theory simple, it is apparently not
   suitable for the investigation of the constituent quark picture.
Thus, we treat the model without integrating the fermion fields.

In addition, it is convenient to introduce a kinetic term in the 
   auxiliary field.
Then the system becomes similar to the Yukawa model, and we recover the 
   Gross-Neveu model by taking the limit $\mu\rightarrow \infty$,
   where $\mu$ is a parameter in the coefficient of the kinetic term.
In this sense, we call this the ``generalized" Gross-Neveu
   model \cite{rf:Pes,rf:Kug}. \ 
Introduction of the auxiliary field and its kinetic term makes the analysis 
   tractable.
The Gross-Neveu model displays some peculiar properties on the LC, 
   which are characteristic of the four-Fermi theories.
For example, the constraint on the nondynamical component of the 
   fermion becomes nonlinear, and it is very difficult to solve.
Also, there is a problem in defining a classical Hamiltonian \cite{rf:Pes}. \ 
While
   these are obstacles for the analysis of the light-front 
   Gross-Neveu model, the generalized Gross-Neveu model avoids these points.

In \S 2, 
   we define the generalized Gross-Neveu model and quantize the model
   following Dirac's method for constrained systems \cite{rf:DirLec}. \ 
This model is a 1+1 dimensional field theory of
   a scalar field $\phi(x)$ and an $N$-component fermion $\psi^a(x)$
   interacting with each other via the Yukawa interaction.
From a canonical argument, we find a zero mode constraint 
   for the scalar field.
This constraint implies that the zero mode of the scalar $\phi(x)$
   should be expressed by fermion's oscillation modes.

In \S 3, 
   the zero-mode constraint is solved using the $ 1/N $ expansion.
It is shown that in the leading order of the $1/N$ expansion, the
   zero mode of $ \phi (x ) $ has a solution whose vacuum expectation value 
   is not zero ($\langle 0|\phi(x)|0\rangle =  \sigma \neq 0 $).
If we choose the nontrivial solution, 
   the discrete chiral symmetry is broken.
For the sake of obtaining a nonzero value of $\sigma$, 
   we must take the continuum limit $L\rightarrow \infty$ very carefully.
We use a damping factor to avoid the infrared divergence 
   which emerges in the continuum limit, and consider its physical
   meaning.

In \S 4, the light-cone Hamiltonian is considered.
It is shown that the vacuum state is an eigenstate of both the light-cone
   momentum $ P^+ $ and the light-cone Hamiltonian $ P^- $ even
   in the case of spontaneous symmetry breaking.
In the large $N$ limit, the obtained light-cone Hamiltonian of the
   Gross-Neveu model has two terms.
The first is the kinetic term of the fermion with mass $\sigma$,
   which is the solution of the zero-mode constraint
   to leading order.
Therefore the excited state of the fermion has a mass gap $\sigma$.
The second is the interaction term of these massive fermions,
   which is written using the remaining operator part of the zero
   mode determined from the zero-mode constraint.
Because we know the mass of the bound state $ \bar{\psi} \psi $ is
   $ 2 \sigma $ in the Gross-Neveu model, it is natural to
   call the fermion's excited state with mass $\sigma$  the ``constituent
   quark."
A conclusion is given in the final section.
%
%
%
\section{The model and its quantization}
\subsection{\it The generalized Gross-Neveu model}

The Gross-Neveu model \cite{rf:GroNev} is a $ 1+1 $ dimensional model
defined by \begin{equation}
{\cal L}_{\rm GN}= ~ \sum_a \bar \psi^a~i \rlap/ \partial~  \psi^a +
           {\lambda_0 \over 2 N} 
             \left(\sum_a \bar \psi^a  \psi^a \right)^2 , \hskip 2cm
              (a=1,2, . . . ,N)
 \label{bc}
\end{equation}
   where $ \psi^a \equiv (~ \psi_R^a ~,~ \psi_L^a ~) $ is an 
   $N$-component massless fermion.
Instead of the Gross-Neveu model,  it is convenient to treat the 
   following $ 1+1 $ dimensional model with the Yukawa 
   interaction \cite{rf:Pes,rf:Kug}
\begin{eqnarray}
{\cal L}= ~ \sum_{a=1}^N \bar \psi^a~i \rlap/ \partial~  \psi^a
            + \frac{N}{2 \mu^2}(\partial_\mu \phi )(\partial^\mu \phi )
            - \frac{N}{2 \lambda_0} \phi^2
            - \sum_{a=1}^N \phi ~ \bar \psi^a  \psi^a  ,
 \label{ba}
\end{eqnarray}
where  ~$\phi$~ is a boson with mass $ \mu / \sqrt{\lambda_0} $.
Both of the lagrangian densities have the discrete chiral symmetry
   $ (~ \psi^a \rightarrow \gamma_5 ~ \psi^a,~ \phi \rightarrow - \phi ~) $
   and a global $ U(N) $ symmetry 
   $ \psi^a \rightarrow U_b^a \psi^b \ (U\in U(N))$.
From tha calculation of the effective potential to leading order in $ 1/N $,
   it was shown that the discrete chiral symmetry is broken
   spontaneously in the large $N$ limit \cite{rf:GroNev,rf:Kug}.                                                          \ 
The Yukawa-like model Eq.\ (\ref{ba}) can be reduced to the Gross-Neveu model
   in the limit $ \mu \rightarrow \infty $.
Therefore, we shall call this
   the ``generalized" Gross-Neveu model in this paper.

The notation we use is as follows.
The light-cone time is $ x^+ \equiv x^0 + x^1 $,
   the light-cone space coordinate is
   $ x^- \equiv x^0 - x^1 $, and the derivatives are
   $ \partial_+ \equiv \partial / \partial x^+ $ and
   $ \partial_- \equiv \partial / \partial x^- $.
The $ \gamma $ -matrices are $ \gamma^0 = \sigma^1 $ and
   $ \gamma^1 = -i \sigma^2 $, and we also define
   $ \gamma^+ \equiv \gamma^0 + \gamma^1 $ and
   $ \gamma^- \equiv \gamma^0 - \gamma^1 $.
Then we have $ \gamma^\mu \partial_\mu = \gamma^0 \partial_0
   + \gamma^1 \partial_1 = \gamma^+ \partial_+ + \gamma^- \partial_- $.

In DLCQ \cite{rf:PauBro}, the fields are put in
   a ``box"
   $ -L \leq x^- \leq L $.
The boson $ \phi ( x ) $ and the fermion field $ \psi^a (x) $ are taken to
   satisfy periodic and antiperiodic boundary conditions, respectively:
\begin{equation}
  \phi( x^+,~x^- =-L~)= \phi( x^+,~x^- =L~),
  \label{bd}
\end{equation}
\begin{equation}
  \psi^a ( x^+,~x^- =-L~)= -\psi^a ( x^+,~x^- =L~).
  \label{be}
\end{equation}
Let us write the boson field $ \phi ( x ) $ as the sum of the zero-mode part
   $ \phi_0 (x^+) $,
\begin{equation}
  \phi_0( x^+) \equiv \frac{1}{2 L} \int_{-L}^{L} d x^-  \phi( x^+,~x^-~),
  \label{bf}
\end{equation}
and the remaining oscillation part $ \varphi( x^+,~x^-~) $:
\begin{equation}
  \phi( x^+,~x^-~)= \phi_0( x^+ ~) + \varphi( x^+,~x^-~).
  \label{bg}
\end{equation}
We treat the zero mode separately from the nonzero modes.
Note that the fermion field does not contain 
   the zero mode because of the antiperiodic
   boundary condition.

\subsection{\it Canonical quantization}

Now let us quantize the generalized Gross-Neveu model using 
   Dirac's method \cite{rf:DirLec} for constrained systems.
Having the $( 2N+2 )$ canonical variables
   $ (~ \phi_0,~ \varphi,~ \psi_R^a~,~ \psi_L^a ~) \equiv \Phi $, 
   we can calculate their conjugate
   momenta $ \Pi_{\Phi} $ through the lagrangian defined by
   $ L \equiv (1/2) \int_{-L}^{L} d x^- \cal{L} $~,
\begin{eqnarray}
  \Pi_{\phi_0} & \equiv & \frac{\delta L}{\delta \dot{\phi_0} } 
                  = 0~ , \label{bi}            \\
  \Pi_{\varphi} & \equiv & \frac{\delta L}{\delta \dot{\varphi} }
           = \frac{N}{\mu^2} ~ \partial_- \varphi ~, \label{bj}     \\
  \Pi_{\psi_R}^a & \equiv & \frac{\delta L}{\delta \dot{\psi_R^a} } 
                  = i \psi_R^{a \dag} ~, \label{bk}            \\
  \Pi_{\psi_L}^a & \equiv & \frac{\delta L}{\delta \dot{\psi_L^a} } = 0 ~,
        \hskip 1cm   ( a= 1,2, \dots ,N )        \label{bl}      
\end{eqnarray}
where $ \dot{\phi_0} $ is the velocity $ \partial_+ \phi_0 $ , and so on.

As is often the case with the quantization of fermions,
   we do not regard Eq.\ (\ref{bk}) as a constraint.
Then we have $( N+2 )$ primary constraints
   $ \theta_j \approx 0 ~ (~ j=1,2, \dots, N+2 ~)$
   \cite{rf:HeiKruSimWer,rf:MasYam},                                                                                                              
\begin{eqnarray}
  \theta_1 & \equiv & \Pi_{\varphi} - \frac{N}{\mu^2} ~ \partial_- \varphi
           \approx 0 ~, \label{bm}            \\
  \theta_2 & \equiv & \Pi_{\phi_0} \approx 0 ~, \label{bn}        \\
  \theta_{a+2} & \equiv &  \Pi_{\psi_L}^a \approx 0 ~.
           \hskip 1cm   ( a= 1,2, \dots ,N )    \label{bo}        
\end{eqnarray}
Since we treat the longitudinal zero mode explicitly, we work with
the following canonical Hamiltonian:
\begin{equation}
  H_c = \int_{-L}^L d x^- \sum_{\Phi} \Pi_{\Phi} \dot{\Phi} - L ~.
  \label{bp}
\end{equation}
The non-vanishing Poisson brackets are \cite{rf:HeiKruSimWer,rf:MasYam}
\begin{eqnarray}
  \{~ \phi_0 ( x^+ )~ ,~ \Pi_{\phi_0} ( x^+ )~ \}^{\rm PB} & = &
                         \frac{1}{2 L}~, \label{bq}          \\
  \{~ \varphi ( x )~ ,~ \Pi_{\varphi} ( y )~ \}_{x^+ =y^+}^{\rm PB} & = &
                       \delta (~x^- - y^- ~) - \frac{1}{2 L}~, \label{br}  \\ 
  \{~ \psi_R^a ( x )~ ,~ \Pi_{\psi_R}^b ( y )~ \}_{x^+ =y^+}^{\rm PB} & = &
           \delta^{ab} \delta (~x^- - y^- ~) ~, \label{bs}  \\ 
  \{~ \psi_L^a ( x )~ ,~ \Pi_{\psi_L}^b ( y )~ \}_{x^+ =y^+}^{\rm PB} & = &
           \delta^{ab} \delta (~x^- - y^- ~) ~. \label{bt}        
\end{eqnarray}
If we introduce the total Hamiltonian $H$,
\begin{eqnarray}
  H & = & H_c + \sum_{j=1}^{N+2} \int_{-L}^L d x^- ~ u_j ~ 
                                        \theta_j \nonumber  \\
    & = & \frac{1}{2} \int_{-L}^L d x^- \left[ \frac{N}{2 \lambda_0}
          ( \phi_0^2 + \varphi^2 ) - 2 i \sum_a \psi_L^{a \dag} \partial_-
              \psi_L^a                        \right.  \nonumber     \\
    &   & \left. \hskip 3cm  + (\phi_0 + \varphi ) \sum_a (~ \psi_R^{a \dag}
          \psi_L^a + \psi_L^{a \dag} \psi_R^a ~) \right] \nonumber  \\
    &  & \hskip 0.5cm + \sum_{j=1}^{N+2} \int_{-L}^L d x^- ~ u_j ~ \theta_j ~,
 \label{bu}
\end{eqnarray}
where $ u_j $ are the Lagrange multipliers, the time development of any
   function $F$ is given by
$
  \dot{F} \approx \{~F~,~H~\}^{\rm PB} ~.
  \label{bv}
$
The consistency condition $ \dot{\theta_1} \approx 0 $ merely determines the
   coefficient $ u_1 $~, whereas the condition $ \dot{\theta_2} \approx 0 $
   gives another constraint ( secondary constraint ) $ \theta_{N+3} $~,
\begin{equation}
  \theta_{N+3} ~\equiv ~\frac{N}{\lambda_0}~ \phi_0 + \frac{1}{2 L} \sum_a
        \int_{-L}^L d x^- (~ \psi_R^{a \dag}
        \psi_L^a + \psi_L^{a \dag} \psi_R^a ~) \approx 0 ~.
  \label{bw}
\end{equation}
This constraint is also derived from longitudinal integration of the 
   Euler-Lagrange equation for $\phi(x)$.

Furthermore, the condition ~$ \dot{\theta}_{a+2} \approx 0 $~ generates the
   following constraints on ~$ \theta_{N+3+a} $~:
\begin{equation}
  \theta_{N+3+a}  ~\equiv ~  2 ~ i ~ \partial_- \psi_L^a
      -  (\phi_0 + \varphi )~
      \psi_R^a \approx 0 ~.   \hskip 1cm   (~ a= 1,2, \dots ,N~) 
  \label{bx}
\end{equation}
This is equivalent to the Euler-Lagrange equation for $\psi_L$.
The conditions ~$ \dot{\theta}_{N+3} \approx 0 $~ and
   ~$ \dot{\theta}_{N+3+a} \approx 0 $~ only determine the coefficients
   $ u_j $~,
   and there no longer appears any constraint.
It is easily seen that the $( 2+N )$ primary constraints
   $ (~\theta_1~,~\theta_2~,~\theta_{2+a}~(~a=1.\dots,N~)~) $ and the $( 1+N )$
   secondary constraints $ (~\theta_{N+3}~,~\theta_{N+3+a}~(~a=1.\dots,N~)~)$
   are all second-class.

 From the constraint Eq.\ (\ref{bw}), the zero mode $ \phi_0 $ is
   expressed by the fermion fields $ \psi_R^a $ and $ \psi_L^a $~.
However $ \psi_L^a $ itself can be written in terms of ~$ \psi_R^a $ ,~$
\phi_0 $~ 
   and ~$ \varphi $~ by use of the second-class constraint Eq.\ (\ref{bx}).
This implies that the zero mode $ \phi_0 $ is not dynamical, and is written
   using other dynamical fields, $ \psi_R^a $ and $ \varphi $.
Therefore Eq.\ (\ref{bw}) essentially corresponds to the zero-mode constraint 
   on the scalar field.
The  $ \phi_0 $ is then called the constrained zero mode,
   and it will be analyzed
   in detail in the next section.

The Dirac bracket \cite{rf:DirLec} is defined by
\begin{equation}
  \{~F~,~G~\}^{\rm DB} = \{~F~,~G~\}^{\rm PB}
        - \sum_{s,s'} \{~F~,~\theta_s ~\}^{\rm PB}
             ( M^{-1} )_{s s'} \{~\theta_{s'}~,~G~\}^{\rm PB} ~.
  \label{bz}
\end{equation}
where $
  M_{s s'} \equiv \{~\theta_s~,~\theta_{s'} ~\}^{\rm PB} \ \ 
      (~ s,s' = 1,2, \dots, 2N+3 ~) ~.$
After some calculations, we obtain the Dirac
   brackets between the unconstrained variables,
\begin{equation}
  \{~\varphi(x)~,~\varphi(y)~\}_{x^+=y^+}^{\rm DB} 
     ~=~ \frac{\mu^2}{N}~ \frac{1}{2}
     \left\{ - \frac{1}{2}~ \epsilon( x^- - y^- ) 
     + \frac{( x^- - y^- ) }{2 L} \right\} ~,
  \label{baa}
\end{equation}
\begin{equation}
  \{~ \psi_R^a ( x )~ ,~ i \psi_R^{b \dag} ( y )~ \}_{x^+ =y^+}^{\rm DB}
     ~=~ \delta^{ab} \delta (~x^- - y^- ~) ~.
  \label{bab}
\end{equation}
Since one can treat the second-class constraints ~$ \theta_j \approx 0 ~
   (~ j=1,2,\dots,2N+3 ~) $ as the strong equations after adopting the Dirac
   brackets, the Hamiltonian Eq.\ (\ref{bu}) becomes
\begin{equation}
  H = \frac{1}{2} \int_{-L}^L d x^- \left[~ \frac{N}{2 \lambda_0}~
          ( \phi_0^2 + \varphi^2 ) + (\phi_0 + \varphi )
          \sum_a  \psi_R^{a \dag} \psi_L^a ~ \right] ~.
  \label{bac}
\end{equation}
Here we find the first merit of treating 
   the generalized Gross-Neveu model.  
In the Gross-Neveu model, the canonically obtained LC Hamiltonian 
   becomes zero, and this is one of the difficulties to be avoided.
In the generalized Gross-Neveu model, however, 
   the Hamiltonian does not vanish  as long as $\mu$ is finite.

In order to quantize the theory, the Dirac brackets are replaced by
   ( anti ) commutation relations divided by
   $ i ~\hbar ~(~ \hbar \equiv 1 ~) $,
\begin{equation}
  [~\varphi(x)~,~\varphi(y)~]_{x^+=y^+} ~=~ \frac{\mu^2}{N}~ \frac{i}{2}
     \left\{ - \frac{1}{2}~ \epsilon( x^- - y^- ) 
     + \frac{( x^- - y^- ) }{2 L} \right\} ~,
  \label{ca}
\end{equation}
\begin{equation}
  \{~ \psi_R^a ( x )~ ,~ \psi_R^{b \dag} ( y )~ \}_{x^+ =y^+}
     ~=~ \delta^{ab} \delta (~x^- - y^- ~) ~.
  \label{cb}
\end{equation}
The second-class constraints ~$ \theta_j \approx 0 ~(~ j=1,2,\dots,2N+3 ~) $
   then become the equations for field operators.

%
%
%
%
\section{Solution to the zero-mode constraint and the dynamical 
   mass generation}
\subsection{\it Solving the constraints}
First, the constraints ~$ \theta_{N+3+a} = 0 ~(~ a=1,2,\dots,N ~) $,  
   Eq.\ (\ref{bx})
   are easily solved, yielding
\begin{equation}
  \psi_L^a(x^+,x^-) = \frac{1}{2 i}~ \frac{1}{2} \int_{-L}^L d y^- ~
     \epsilon (x^- - y^-) \{ \phi_0(x^+) + \varphi(x^+,y^-) \} ~
     \psi_R^a(x^+,y^-) ~,
  \label{cc}
\end{equation}
where we have discarded the constant of integration \cite{rf:PauBro}. \ 
This is a desirable result because
   in the Gross-Neveu model, the constraint on the nondynamical 
   component of the
   fermion is nonlinear and difficult to solve.
Inclusion of the scalar field makes the constraint easier to solve.
This is the second motivation for us to treat the 
   generalized Gross-Neveu model.

Substituting Eq.\ (\ref{cc}) into Eq.\ (\ref{bw}), 
   we obtain the zero-mode constraint.
In passing from a classical theory to a quantum theory,
   one must prescribe the ordering of the
   field operators and the regularization.
Since the zero-mode operator $ \phi_0 $ is a function of ~$ \psi_R^a $~,
   $ \psi_R^{a \dag}$ and $ \varphi $~, ~$ \phi_0 $ does not commute with
   these operators,~ and there arises the
   problem of ordering.
Here, we simply take a particular ordering of them.
This problem will be discussed elsewhere in similar
   models \cite{rf:ItaMae}. \ 
We take the following ordering in the
constraint relation: %
\begin{eqnarray}
  \phi_0 & = & \frac{\lambda_0}{N}~ \frac{i}{4}~ \frac{1}{2L}~
     \sum_{a=1}^N \int_{-L}^L d x^- \int_{-L}^L d y^- ~\epsilon (x^- - y^-)
               \nonumber    \\
  & & \hskip 2cm \times \{ \phi_0 + \varphi(y^-) \}
     ~\psi_R^{a \dag}(x^-) \psi_R^a(y^-) + ~ \rm{h.c.}  \nonumber    \\
        & = & \left\{ 
             \frac{\lambda_0}{N}~ \frac{i}{4}~ \frac{1}{2L}
             \sum_{a=1}^N \int_{-L}^L d x^- \int_{-L}^L d y^- ~
             \epsilon (x^- -y^-)  \phi_0  ~\psi_R^{a \dag}(x^-)
             \psi_R^a(y^-) + \rm{h.c.}  \right\}   \nonumber    \\
        &   & \hskip 0.01cm + \left\{ 
             \frac{\lambda_0}{N}~ \frac{i}{4}~ \frac{1}{2L}
             \sum_{a=1}^N \int_{-L}^L d x^- \int_{-L}^L d y^- ~
             \epsilon (x^- -y^-)  \varphi(y^-)  ~\psi_R^{a \dag}(x^-)
             \psi_R^a(y^-) + \rm{h.c.}  \right\}~,   \nonumber   \\
       &    &
  \label{cd}
\end{eqnarray}
where~ h.c. represents the Hermitian conjugate.
The constraint relation Eq.\ (\ref{cd}) determines the zero mode $ \phi_0 $~,
   whose vacuum expectation value determines whether the discrete chiral
symmetry
   is broken or not.

Here we expand the field $ \varphi(x) $ at $ x^+=0 $ as a Fourier integral over
   plane waves to obey the periodic boundary condition Eq.\ (\ref{bd}) ,
\begin{equation}
  \varphi(0,x^-) = \frac{\mu}{\sqrt{N}}~ \frac{1}{\sqrt{4 \pi}}~
     \sum_{n=1,2, \dots}^{\infty} \frac{1}{\sqrt{n}} 
     \left( a_n ~ e^{- i k_n^+ x^- /2 } + a_n^{\dag} ~ e^{\, i k_n^+ x^- /2 }
         \right) ~,
  \label{ce}
\end{equation}
where the discrete light-cone momentum $ k_n^+ $ is
\begin{equation}
  k_n^+ = \frac{2 \pi}{L} n ~.
  \label{cf}
\end{equation}
The fermion field $ \psi_R^a (x) $ is also expanded at $ x^+=0 $, giving
\begin{equation}
  \psi_R^a(0,x^-) = \frac{1}{ \sqrt{2L}}~
    \sum_{n= \frac{1}{2}, \frac{3}{2}, \dots}^{\infty}
    \left( b_n^{\, a} ~ e^{- i k_n^+ x^- /2 } + d_n^{\, a \dag}~
     e^{\, i k_n^+ x^- /2 } \right) ~,
  \label{cg}
\end{equation}
where $n$ is a half-integer to satisfy the antiperiodic boundary condition
    Eq.\ (\ref{be}) .
The (anti) commutation relations
\begin{eqnarray}
  & [~ a_n~,~a_m^{\dag} ~] = \delta_{nm} ~,  \nonumber   \\
  & \{~b_n^{\, a} ~, ~b_m^{\, b \dag} ~\} = \delta^{ab} \delta_{nm} \: ,
  \hskip 1cm
    \{~d_n^{\, a} ~, ~d_m^{\, b \dag} ~\} = \delta^{ab} \delta_{nm} \: ,
  \label{ch}
\end{eqnarray}
follow from Eqs.\ (\ref{ca}) and (\ref{cb}) .
The vacuum state $ |0 \rangle $ is defined so as to satisfy
   $ a_n | 0 \rangle =0 $,
   $ b_n^{\, a} | 0 \rangle =0 $ and $ d_n^{\, a} | 0 \rangle =0 $.
The Fock space is constructed by operating the creation operators
   $ a_{n}^{\dag} $~,
   $ b_n^{\, a \dag} $ and $ d_n^{\, a \dag} $ on $ |0 \rangle $ .
Since all these quanta have the positive light-cone momentum $ k_n^+ >0 $~,
   the state $ | 0 \rangle $ is the only one which has zero light-cone momentum
   $ k^+=0 $~, and therefore the vacuum is trivial.

Before solving the zero-mode constraint Eq.\ (\ref{cd}) , we shall write
the zero
   mode $ \phi_0 $ as a sum of the vacuum expectation value
   $ \langle 0 | \phi_0 | 0 \rangle  \equiv \sigma $ and the remaining
   operator part
   $ ( \phi_0 )_{\rm oper} $~,
\begin{eqnarray}
  \phi_0 & = & \langle 0 | \phi_0 | 0 \rangle +~ ( \phi_0 )_{\rm oper}
                               \nonumber   \\
         & = & \hskip 0.6cm  \sigma \hskip 0.6cm   + ( \phi_0 )_{\rm oper} ~,
  \label{ci}
\end{eqnarray}
where $ ( \phi_0 )_{\rm oper} $ is normal ordered with respect to the Fock 
   vacuum  $|0\rangle$.
If we find a solution such as $ \sigma \neq 0 $, 
   the discrete chiral symmetry is broken spontaneously.

Now, we show that the zero-mode constraint Eq.\ (\ref{cd})  does have a
   nontrivial solution $ \sigma \neq 0 $ in the large $N$ limit.
Putting the mode expansion Eq.\ (\ref{cg})  into the first term of the
   right-hand side of Eq.\ (\ref{cd}), and using the integral
\begin{equation}
  \frac{i}{4 \pi} \int_{-\pi}^{\pi} d \xi \int_{-\pi}^{\pi} d \xi'
     ~ \epsilon(\xi - \xi') ~ e^{\, i (n \xi + m \xi') }~ =~ - \frac{1}{n}~
        \delta_{n,-m} \; ,   \nonumber
\end{equation}
\begin{displaymath}
  \hskip 4cm (~ n~,~m= \dots , -\frac{3}{2}~ ,~ -\frac{1}{2} ~,~
      \frac{1}{2} ~,~
       \frac{3}{2} ~,~\dots ~)
  \label{cl}
\end{displaymath}
we find that the zero-mode constraint takes the 
\begin{eqnarray}
  \phi_0 & = & \left\{
    -\frac{\lambda_0}{4 \pi}~ \frac{1}{N} \sum_a
    \sum_{n= \frac{1}{2} ,\frac{3}{2} ,\dots }^{\infty} \phi_0 ~
    ( \frac{ \triangle k_n^+ }{ k_n^+}  )~
    (~ b_n^{\, a \dag} b_n^{\,a}
     - d_n^{\,a} d_n^{\, a \dag} ~) + \rm{h.c.}  \right\}  \nonumber    \\
   &  &  + \left\{ \hskip 0.5cm {\rm the ~~term~~ involving~~ }~ \varphi(y^-)
                          \hskip 0.5cm  \right\} ,  
  \label{cma}
\end{eqnarray}
where ~$ k_n^+ = 2 \pi n /L $~ is the light-cone momentum of the quanta
   $ b_n^{\,a} $ or $ d_n^{\,a} $~, and~
   $ \triangle  k_n^+ \equiv  k_{n+1}^+ - k_n^+ = 2 \pi /L $.
The vacuum expectation value of
    Eq.\ (\ref{cma}) becomes
\begin{eqnarray}
  \langle 0| \phi_0 |0 \rangle 
     &   = &
      -\frac{\lambda_0}{4 \pi}~ \frac{1}{N} \sum_a
      \sum_{n= \frac{1}{2} }^{\infty}~
      \langle 0 | ~\phi_0 ~( \frac{ \triangle k_n^+ } { k_n^+ }  )~
      (~ b_n^{\, a \dag} b_n^{\,a}
       - d_n^{\,a} d_n^{\, a \dag} ~) ~|0 \rangle
       + \rm{h.c.}                        \nonumber    \\
     & = & \frac{\lambda_0}{2 \pi}~ \sigma
            \sum_{n= \frac{1}{2} }^{\infty}
            \frac{  \triangle k_n^+ } { k_n^+ } ~~,
  \label{cna}
\end{eqnarray}
where we have used $  \langle 0| \varphi |0 \rangle = 0 $.
Although this equation is obtained without taking the $1/N$ expansion,
   this is due to our particular operator ordering.
If we assumed a different ordering, we would have to perform the $1/N$ expansion
   to obtain the same result as Eq.\ (\ref{cna}).
Since a nonzero solution $\sigma\neq 0$ corresponds to the fermion mass,
   we use the dispersion relation 
\begin{equation}
  k_n^- = \frac{\sigma^2}{k_n^+} ~.
  \label{cj}
\end{equation}
Now let us introduce the ultraviolet cutoff $\Lambda'$ to 
   regulate the divergent sum $ \sum_{n= \frac{1}{2} }^{\infty}
            ~ \triangle k_n^+ /  k_n^+  $,
\begin{equation}
  k_n^+ ~ \leq ~ \frac{ 2 \pi }{ L }~ \left ( \frac {R}{2} \right )
      \equiv \Lambda ' ~,
  \label{cza}
\end{equation}
where $R$ is a large odd number, and $ \Lambda ' $ is the ultraviolet cutoff
   parameter which has dimensions of mass.
Then we have
\begin{equation}
  \sum_{n= \frac{1}{2} }^{R/2}~\frac { \triangle k_n^+ } { k_n^+ }
  ~ \simeq ~ \log \frac{R}{2} + \log 4 e^\gamma ~
  = ~ \log { L \Lambda'} + \log \frac{2 e^\gamma}{\pi} ~,
  \label{czb}
\end{equation}
where $\gamma$ is Euler's constant.
Unfortunately, the summation has an infrared divergence in the limit 
  $ L \rightarrow \infty $ in addition to the ultraviolet
  divergence \cite{rf:KojSakSak}. \ 
This is because two independent parameters $L$ and $\Lambda'$ are introduced
   as
\begin{equation}
  \frac{ 2 \pi }{ L } \left ( \frac {1}{2} \right )
           ~ \leq ~ k_n^+  ~ \leq ~  \Lambda' ~.
  \label{czd}
\end{equation}
Physical quantities should not depend on the artificial parameter $L$.
( The ultraviolet cutoff $ \Lambda' $ will be renormalized into
   the coupling constant. )
To avoid the infrared divergence, Kojima et al. \cite{rf:KojSakSak}
   introduced quantization surfaces interpolating between the light-cone
   surface and the spatial surface, and defined physical
   quantities unambiguously.
They found that the procedure to avoid the infrared divergence is
   important to reproduce the double well effective potential (spontaneous
   symmetry breaking) in the light-front Gross-Neveu model.

Here, in order to avoid the infrared divergence in Eq.\ (\ref{czb}),
   we will introduce a damping factor $ f( k^+ ) $.
Let us use the ultraviolet cutoff $ \Lambda' $
   in Eq.\ (\ref{cza}) to determine the damping factor.
The function $ f( k^+ ) $ satisfies
\begin{equation}
   \lim_{\Lambda' \rightarrow \infty} f( k^+ ) =1
  \label{czo}
\end{equation}
and
\begin{equation}
    f( k^+ ) \simeq \left\{ \begin{array}{ll}
       0   &  \hskip 1cm ( k^+ \simeq  0 )           \\
       1 ~.  &  \hskip 1cm ( k^+ \simeq  \Lambda' )
     \end{array}
    \right.  
  \label{czp}
\end{equation}
There will be many functions having the above properties;
   a simple one is of the heat-kernel type,
\begin{equation}
   f( k^+ ) = e^{- k^-/\Lambda' }
            = e^{- \sigma^2 / ( \Lambda' k^+ ) }~,
  \label{czq}
\end{equation}
which effectively introduces an infrared cutoff 
   $ k^+ \sim \sigma^2 / \Lambda' $
   depending on the fermion mass $\sigma$.

With this damping factor, the summation in the large $N$ limit
\begin{equation}
  \sum_{n= \frac{1}{2} }^{R/2}~\frac { \triangle k_n^+ } { k_n^+ }~ f( k^+_n )
   = \sum_{n= \frac{1}{2} }^{R/2}~\frac { \triangle k_n^+ } { k_n^+ }~
     e^{- \sigma^2 / ( \Lambda' k_n^+ ) }
  \label{czr}
\end{equation}
does not exhibit an infrared divergence, and we can take
   $ L \rightarrow \infty $ :
\begin{equation}
  \sum_{n= \frac{1}{2} }^{R/2}~\frac { \triangle k_n^+ } { k_n^+ }~
     e^{- \sigma^2 / ( \Lambda' k_n^+ ) }
   \longrightarrow
   \int_0^{\Lambda'} d k^+ \frac{1}{k^+}~ e^{- \sigma^2 / ( \Lambda' k^+ )}
   = \log \frac{ \Lambda^{\prime 2} }{\sigma^2} - \gamma
       + O (  \frac{\sigma^2}{ \Lambda^{\prime 2} } ).
  \label{czs}
\end{equation}
As we will see below, the integral 
   $  \int_0^{\Lambda'} d k^+ ( 1/ k^+ )  f( k^+ ) $ 
   corresponds to the loop integral of the fermion with mass $\sigma$ in
   the equal-time quantization formalism.
Therefore it is natural for the term 
   $ \log \Lambda^{\prime 2} / \sigma^2 $
   to appear in Eq.\ (\ref{czs}) when the divergent sum
   $ \sum_{n= \frac{1}{2} }^{\infty}~ \triangle k_n^+ /  k_n^+  $
   in DLCQ is regularized properly.

Another choice of the function $ f( k^+ ) $ would give the same
   physical results.
For example, if we choose the damping factor 
   $ f( k^+ ) = k^+ / ( k^+ + \sigma^2/\Lambda' )$ ,
   the physical results are the same as above.

Eventually, the zero-mode constraint  Eq.\ (\ref{cna}) in the large $N$ limit
   becomes
\begin{eqnarray}
     0
     &   = &
      \sigma \left[ \frac{1}{\lambda_0} + \frac{1}{2\pi} 
      \left\{ \log \frac{\sigma^2}{\Lambda^{\prime 2} }
                              + \gamma \right\}  \right]
                                                        \nonumber    \\
     & = & 
      \sigma \left[ \frac{1}{\lambda_0} + \frac{1}{2\pi} 
      \log \frac{\sigma^2}{\Lambda^2}
      \right]  ~~,
  \label{czi}
\end{eqnarray}
where $ \Lambda \equiv e^{- \gamma/2} \Lambda' $ .
This is nothing but the gap equation.
We define the renormalization coupling constant $\lambda_R (M) $ as
\begin{equation}
  \frac{1}{\lambda_R (M) } ~\equiv ~ \frac{1}{\lambda_0}
      + \frac{1}{2\pi}  \log \frac{M^2}{\Lambda^2} 
      + \frac{1}{\pi} ~,
  \label{cs}
\end{equation}
where $M$ is a positive finite quantity corresponding 
   to the renormalization point.
By use of $ \lambda_R $~, Eq.\ (\ref{czi}) is rewritten as
\begin{equation}
  \sigma \left[ \frac{1}{\lambda_R (M) } + \frac{1}{2\pi} 
      \log (\frac{\sigma^2}{M^2}) - \frac{1}{\pi} \right] = 0 ~.
  \label{ct}
\end{equation}
This equation indeed has a nontrivial solution,
\begin{equation}
  \sigma = M e^{(1-\pi/ \lambda_R )} > 0 ~.
  \label{cu}
\end{equation}

Hence, the zero-mode constraint has the nontrivial
   solution ~$ \sigma = \langle 0 | \phi_0 |0 \rangle ~ > 0 $,
   and therefore, the discrete chiral
   symmetry is broken spontaneously and the massless fermion obtains mass
   $ \sigma $ dynamically.
The fact that the nontrivial solution $ \sigma\neq 0$ does not depend on the
   parameter $ \mu $ implies that the spontaneous chiral symmetry
   breaking also occurs in the Gross-Neveu model.
In deriving the nontrivial solution Eq.\ (\ref{cu}), it is essential to treat
   the infrared divergence in Eq.\ (\ref{czb}) correctly in DLCQ.

There is one thing on which we should comment here.
Certainly, we obtain a nontrivial solution $\sigma\neq 0$,
   but Eq.\ (\ref{ct}) also has a trivial (symmetric) solution $\sigma=0$.
The zero-mode constraint itself does not reveal 
   which solution is realized in reality.
For this purpose, we must calculate the vacuum energy.
This point is discussed in the next section.

\subsection{\it The physical meaning of the infrared regularization}

Here we explain the physical meaning of our infrared regularization
   method (i.e., the damping factor $ f(k^+) $ ).
   
Before we explain the meaning of our infrared regularization method, it
   would be a help if we knew the other derivation of Eq.\ (\ref{cna}), 
   that is, the mean field approximation \cite{rf:DieHeiSchWer}. \ 
Possibly, this is the simplest way to understand the origin of
   the infrared divergence in Eq.\ (\ref{cna}).
In order to simplify, we consider the Gross-Neveu model Eq.\ (1).
Let us perform an approximation which linearizes the Euler-Lagrange
   equations:
\begin{eqnarray}
   2 ~ i~\partial_+ \psi_R^a &=& \sigma ~ \psi_L^a  ~ ,  \\ 
   2 ~ i~\partial_- \psi_L^a &=& \sigma ~ \psi_R^a  ~,
\end{eqnarray}
where  $ \sigma $ is the effective mass of the fermion.
If we evaluate the mean field equation $\sigma = - (\lambda_0/N)
                \langle 0| \bar{\psi} \psi| 0\rangle$
   by using the above (massive) fermion, 
   we find that it reproduces  Eq.\ (\ref{cna}),
\begin{eqnarray}
   \sigma &=& - \frac{\lambda_0}{N} ~
                \langle 0| \bar{\psi} \psi| 0\rangle  \nonumber \\
          &=& - \frac{\lambda_0}{N} ~ \sum_a ~
                \langle 0| \psi_R^{a \dagger} \psi_L^a 
                + \psi_L^{a \dagger} \psi_R^a | 0\rangle  \nonumber \\
          &=& \frac{\lambda_0}{2 \pi} \sigma \sum_{n=1/2}^\infty
                \frac{ \triangle k_n^+}{k_n^+} .
\end{eqnarray}
What has to be noted in this calculation is that the momentum
   sum came from just the vacuum expectation value of
   $\bar{\psi}\psi$ ~: 
\begin{eqnarray} 
   \langle 0| \bar{\psi}\psi| 0\rangle
   &=&  \frac{\sigma}{2\pi} \sum_a \sum_n ~
         \left( \frac{ \triangle k_n^+}{k_n^+} \right) ~
        \langle 0| ( b_n^{a \dagger} b_n^a - d_n^a d_n^{a \dagger} )
        | 0\rangle                                       \nonumber \\
   &=& - \frac{N}{2\pi} \sigma \sum_n \frac{ \triangle k_n^+}{k_n^+} 
                                                           \nonumber \\
   & \longrightarrow & 
       - \frac{N}{2 \pi} \sigma \int_0^\infty \frac{dk^+}{k^+} .
\end{eqnarray} 
This observation provides an understanding of the meaning of our infrared
regularization scheme.

We introduced the infrared cutoff so that it depended on the ultraviolet 
   cutoff.
This was crucial to obtain a nontrivial solution $\sigma\neq 0$.
Such infrared regularization has been used in several
   works \cite{rf:DieHeiSchWer,rf:HarVar},
   but in these works its physical meaning is obscure and its necessity is not
   discussed.
Thus we explain the physical meaning of our infrared
   regularization method in detail.

To this end, we must remember that the light-cone formalism suffers
   from a loss of mass information from the Green's
   functions \cite{rf:NakYam}.
This is a common problem in LC field
   theories, and our infrared regularization 
   is understood as one of the prescriptions
   for obtaining a correct result even in the LC field theories.
The mean field approximation taught us that the divergent integral in effect 
   came from $\langle 0| \bar{\psi}\psi| 0\rangle$.
Let us show in the very elementary case, the free massive fermion,
   that we must include the mass information to obtain a correct result.

In the equal-time formulation, the term causing the divergence is
   simply given by
\begin{equation}
   \langle 0|\bar{\psi}(x) \psi (y) | 0\rangle 
    =-{\rm Tr} S_F(y-x)\ ,
\end{equation} 
\begin{equation} 
   S_F(x)=\int \frac{d^2 k}{i(2\pi)^2}
           \frac{e^{-ikx}}{m-k\!\!\!/-i\epsilon} ,
\end{equation} 
where the trace is for the color and spinor indices.  
There is a logarithmic divergence here, and we need a cutoff to regulate it.  
So we introduce the cutoff function $F(k,\Lambda)$:
\begin{equation} 
   \langle 0|\bar{\psi}(x) \psi(x)| 0\rangle
     = 2Nm\int\frac{d^2k}{i(2\pi)^2}\frac{1}{m^2-k^2-i\epsilon}
       F(k,\Lambda).  \label{reg_int}
\end{equation} 
If we evaluate this in Euclidean space and use the covariant cutoff
$F(k,\Lambda)=\theta(\Lambda^2-k_{\rm E}^2)$, we obtain 
\begin{equation} 
   \langle 0|\bar{\psi}\psi| 0\rangle 
    = -\frac{N}{2\pi}\ m\ {\rm ln}\frac{\Lambda^2+m^2}{m^2}.  
\end{equation} 
Next let us evaluate Eq.\ (\ref{reg_int}) in light-cone variables.  
This is achieved simply by changing variables.
The $k^-$ integration can be easily performed as 
\begin{eqnarray} 
   \langle 0| \bar{\psi}(x)\psi(x)| 0\rangle
   &=& 2N\ m\int\frac{d^2k}{i(2\pi)^2}
        \frac{1}{m^2-k^2-i\epsilon}F(k,\Lambda)    \nonumber\\
   &=&-2Nm\int\frac{dk^+ dk^-}{2i(2\pi)^2}
       \frac{F(k,\Lambda)}{k^+ \left(k^- -\frac{m^2}{k^+} +i\epsilon\ 
       {\rm sgn}(k^+)\right)}                      \nonumber\\
   &=&-\frac{N}{2\pi}m\int_0^\infty\frac{dk^+}{k^+} 
       F(k^-=m^2/k^+,k^+,\Lambda).   \label{ETLC} 
\end{eqnarray} 
This calculation instructs us that
   {\it if we use light-cone variables, the mass-dependence enters 
   the
   integration  only through the cutoff functions}.
On the other hand, as we saw\footnote{ After the mean field
   approximation, the model became that of a free massive fermion.} 
   in the mean field calculation,
   the LC quantization gives 
\begin{equation} 
   \langle 0| \bar{\psi} (x) \psi (x) | 0\rangle
     =-\frac{N}{2\pi}\ m\ \int_0^\infty\frac{dk^+}{k^+}.  
\end{equation} 
Apparently we do not have any mass dependence in the integration.
Comparing this with the result of the equal-time quantization Eq.\
(\ref{ETLC}), 
   it is evident that we cannot
   obtain the correct result even in the free theory unless we include
   any mass information into the integration as the regularization.

The loss of mass information in the two-point function has already been
   pointed out by Nakanishi and Yamawaki as the most fundamental
   problem of the light-front field theories \cite{rf:NakYam}. \ 
In the scalar theory, they compared the two point function derived
   from the general argument of the equal-time formulation and that
   from the LC quantization.
 From the spectrum representation, we obtain
\begin{equation} 
   \langle 0|\phi(x^+,{\bf x})\phi(x^+,{\bf y})|0\rangle 
     =\int_0^\infty d\kappa^2 \rho(\kappa^2)
       \Delta^{(+)}(0,{\bf x}-{\bf y};\kappa^2) ~,
\end{equation} 
where $\rho(\kappa^2)$ is the spectral function and, $\Delta^{(+)}$ is 
   one of the invariant delta functions.
On the other hand, the conventional LC quantization leads to
\begin{equation} 
   \langle 0|\phi(x^+,{\bf x})\phi(x^+,{\bf y})|0\rangle 
     = Z^{-1} \int \frac{dp^+d^2p_\perp}{(2\pi)^3}\
       \frac{\theta(p^+)}{2p^+}\ e^{-i{\bf p(x-y)}}\ , 
\end{equation} 
where $Z^{-1}=\int_0^\infty d\kappa^2 \rho(\kappa^2)$.
The discrepancy is quite clear.
Namely, the result of the LC quantization has no
   mass dependence.
To remedy this, Nakanishi and Yamawaki included the mass information as
   the regularization of the divergent integral.
The problem with which we have been faced in the (generalized) Gross-Neveu
   model is essentially the same.
The LC field theories always have this
   kind of problem, and we must bear it in mind.

Now the meaning of our infrared regularization is clear.
The inclusion of mass information is necessary for the correct
   evaluation of $\langle 0|\bar\psi\psi|0\rangle$ , and our
   infrared regularization gives one of prescription for this.
As is evident from Eq.\ (\ref{ETLC}), if we choose the same cutoff
   function as in the equal-time calculation, then  the result
   will of course become the same value \cite{rf:DieHeiSchWer}. \ 
However this choice seems very artificial because it deviates from the LC
   theories.
If we did not know the cutoff method in the equal-time
   formulation, we could not attain the correct evaluation.

It is also evident that we should consider the ``inclusion of
   the mass information" and the ``use of observed mass" separately.
The loss of mass information is a common problem 
   in the LC field theories and it has nothing to do with the spontaneous
    symmetry breaking.
The essential step to succeed in obtaining the nontrivial solution 
   is to use
   the observed mass $\sigma$ in regularizing the
   integral.

In the next section, the case of the Gross-Neveu model
   $ ( \mu \rightarrow \infty ) $ is studied.
We calculate the light-cone Hamiltonian and discuss the constituent
   quark picture.
%
%
%
%
\section{Constituent picture in the Gross-Neveu model}

In the equal-time quantization, the vacuum structure of the theory with
   spontaneous symmetry breaking is very complicated, and thus it
   is difficult to construct excited states.
On the other hand, in the discretized light-cone quantization, the vacuum is
   quite simple, and the Fock space is easily obtained.

In this section, we calculate the light-cone Hamiltonian of the
   Gross-Neveu model in the large $N$ limit and discuss the constituent
   quark picture.

The lagrangian Eq.\ (\ref{ba}) is invariant under translation,
   and the total momentum $ P^+ $ is conserved \cite{rf:PauBro}:
\begin{eqnarray}
  P^+ & = & \int_{-L}^L d ( \frac{1}{2} x^- ) J^{++} 
            = \frac{1}{2} \int_{-L}^L d x^- \left[ 2 i
             \sum_{a=1}^{N} \bar \psi^a \gamma^+ \partial_- \psi^a
              + 4~ \frac{N}{\mu^2}~ \partial_- \phi~ \partial_- \phi  
                    \right]                       \nonumber  \\
      & = & \frac{1}{2} \int_{-L}^L d x^- \left[ 4 i
             \sum_{a=1}^{N} \psi_R^{a \dag} \partial_- \psi_R^a
               + 4~ \frac{N}{\mu^2}~ \partial_- \varphi~ \partial_- \varphi  
                                      \right] ~.
  \label{da}
\end{eqnarray}
Note that there is no zero-mode contribution here.
 From the mode expansion Eq.\ (\ref{ce}) and Eq.\ (\ref{cg}),
   the normal ordered $ P^+ $ is written as
\begin{equation}
  P^+ = \sum_{a=1}^{N} ~ \sum_{n= \frac{1}{2} }^{\infty} k_n^+
        ( ~ b_n^{\, a \dag} b_n^{\,a} + d_n^{\, a \dag} d_n^{\,a} ~)
         + \sum_{n= 1 }^{\infty} k_n^+ a_n^{\dag} a_n ~.
  \label{db}
\end{equation}
Since we defined the lagrangian $L$ by
   $ L \equiv (1/2) \int {\cal L} d x^- $, the (classical) total energy
   $ P^- $ equals $ 2H $, Eq.\ (\ref{bac}) \cite{rf:HeiKruSimWer},
\begin{equation}
  P^- = 2 H = \int_{-L}^L d x^- \left[~ \frac{N}{2 \lambda_0}~
          ( \phi_0^2 + \varphi^2 )~+~ ( \phi_0 + \varphi )
         \sum_a  \psi_R^{a \dag} \psi_L^a ~ \right] ~.
  \label{dd}
\end{equation}
As discussed in \S 3, we should prescribe the ordering of operators
  in the quantum total energy $ P^- $.
We take the following ordering
   in this paper:
\begin{eqnarray}
  P^- & = & \int_{-L}^L d x^-  \frac{N}{2 \lambda_0}~
          ( \phi_0^2 + \varphi^2 )           \nonumber  \\
    & & \hskip 0.1 cm + ~ \frac{1}{4~i}~
     \sum_{a=1}^N \int_{-L}^L d x^- \int_{-L}^L d y^- ~\epsilon (x^- - y^-)
               \nonumber    \\
  & & \hskip 0.7cm \times \left[ \; \; \frac{1}{2} \left\{ \phi_0^2
     ~\psi_R^{a \dag}(x^-) \psi_R^a(y^-)
      ~+~ \psi_R^{a \dag}(x^-) \psi_R^a(y^-) \phi_0^2  \right\}
                                              \right.         \nonumber  \\
  & & \hskip 1.5cm
        +~ \phi_0 ~\psi_R^{a \dag}(x^-) \psi_R^a(y^-) \varphi(y^-)
        +~ \varphi(x^-) ~\psi_R^{a \dag}(x^-) \psi_R^a(y^-) \phi_0 
                                                              \nonumber  \\
  & & \hskip 1.5cm \left.
        +~ \varphi(x^-) ~\psi_R^{a \dag}(x^-) \psi_R^a(y^-) \varphi(y^-)
                               \;  \;  \right]  \: .
  \label{dfa}
\end{eqnarray}
Here, Eq.\ (\ref{cc}) has been used.

Here let us show that the vacuum state $ |0 \rangle $ defined in \S 3
   is an eigenstate of the total energy $ P^- $,  Eq.\ (\ref{dfa}).
In the $ 1+1 $ dimensional light-cone Yukawa theory with no color
   $ (N=1) $, it was shown by Pauli and Brodsky \cite{rf:PauBro}
   that the Fock vacuum
   state  $ |0 \rangle $ is an eigenstate of the light-cone Hamiltonian.
The essential difference between their Hamiltonian and ours Eq.\ (\ref{dfa}) 
   is the existence of the zero mode $ \phi_0 $.
The zero-mode operator ~$ \phi_0 $~ does not change the momentum of states.
Thus the state ~$ \phi_0 |0 \rangle $ has zero momentum.
Since $~|0 \rangle $ is the only state which has zero momentum,
   $ \phi_0 |0 \rangle $ must be a c-number times $|0 \rangle $.
So we can write it as
\begin{equation}
  \phi_0 | 0 \rangle = \sigma | 0 \rangle .
  \label{dzs}
\end{equation}
In other words, the operator part of
   $\phi_0 \ (\phi_0=\sigma+(\phi_0)_{\rm oper})$
   does not include terms  such as $a_na_m,\ b_nb_m$ or 
   $a_n^\dagger a_m^\dagger$,
   but it does contain at least one 
   annihilation or one creation operator, 
   and therefore $\phi_0|0\rangle=\sigma|0\rangle$ and 
   $\langle 0|\phi_0=\langle 0|\sigma$. 
 From this equation, we find that the vacuum state $|0 \rangle $ is
   an eigenstate of $P^-$,  Eq.\ (\ref{dfa})  .
This character and the triviality of the vacuum state $ | 0 \rangle $ 
   are very important when we attempt to construct the constituent quark
   picture.

 From this point, let us consider the symmetry broken phase
   $ \sigma = \langle 0| \phi_0 |0 \rangle \neq 0 $, and the case of the
   Gross-Neveu model, which is obtained when we take the limit 
   $ \mu \rightarrow \infty $ as discussed in \S 2.
In this case, the boson quantum $ a_n^{\dag} $ does not appear in the
   Fock space, so we can ignore the boson term in Eq.\ (\ref{db}):
\begin{equation}
  P^+ = \sum_{a=1}^{N} ~ \sum_{n= \frac{1}{2} }^{\infty} k_n^+
        ( ~ b_n^{\, a \dag} b_n^{\,a} + d_n^{\, a \dag} d_n^{\,a} ~)~.
  \label{dzq}
\end{equation}
Up to leading order in $ 1/N $, the total energy $ P^- $ is given by
\begin{eqnarray}
  P^- & = & \int_{-L}^L d x^-  \frac{N}{2 \lambda_0}~
          \phi_0^2      \nonumber  \\
    & & \hskip 0.5 cm + ~ \frac{1}{4~i}~
     \sum_{a=1}^N \int_{-L}^L d x^- \int_{-L}^L d y^- ~\epsilon (x^- - y^-)
               \nonumber    \\
  & & \hskip 2cm \times \left[ \frac{1}{2} \left\{ \phi_0^2
     ~\psi_R^{a \dag}(x^-) \psi_R^a(y^-)
      ~+~ \psi_R^{a \dag}(x^-) \psi_R^a(y^-) \phi_0^2
         \right\} \right]  \: ,
  \label{df}
\end{eqnarray}
because $ O( \phi_0 ) \sim 1 $ and $ O( \varphi ) \sim 1/ {\sqrt{N}}   $.
 From the mode expansion Eq.\ (\ref{cg})  ,
\begin{eqnarray}
  \frac{P^-}{2L} & = &  \frac{N}{2 \lambda_0}~
          \phi_0^2      \nonumber  \\
    & & \hskip 0.5 cm + ~ \frac{1}{2L} \left[~ \frac{1}{2}~ \phi_0^2
         \sum_{a=1}^{N} ~ \sum_{n= \frac{1}{2} }^{\infty}
          \frac{1}{k_n^+}~
        ( ~ b_n^{\, a \dag} b_n^{\,a} -  d_n^{\,a} d_n^{\, a \dag}~)
        + \rm{h.c.}~ \right]   \nonumber    \\
    & = &  \frac{N}{2}~ \phi_0^2 \left[
            \frac{1}{ \lambda_0}~
           -  \frac{1}{L}~ \sum_{n= \frac{1}{2} }^{\infty}
            \frac{1}{k_n^+}   \right]              \nonumber    \\
    & & \hskip 0.5 cm +  \frac{1}{2L} \left[~ \frac{1}{2}~ \phi_0^2
         \sum_{a=1}^{N} ~ \sum_{n= \frac{1}{2} }^{\infty}
          \frac{1}{k_n^+}~
        ( ~ b_n^{\, a \dag} b_n^{\,a} + d_n^{\, a \dag} d_n^{\,a}  ~)
        + \rm{h.c.}~ \right]                       \nonumber    \\
    & = &  \frac{1}{2L} \left[~ \frac{1}{2}~ \phi_0^2
         \sum_{a=1}^{N} ~ \sum_{n= \frac{1}{2} }^{\infty}
          \frac{1}{k_n^+}~
        ( ~ b_n^{\, a \dag} b_n^{\,a} + d_n^{\, a \dag} d_n^{\,a}  ~)
        + \rm{h.c.}~ \right] ~,
  \label{dg}
\end{eqnarray}
where Eq.\ (\ref{cna}) has been used in the last line.
The zero mode $ \phi_0 $ is a nontrivial operator solution of the
   zero-mode constraint.

We can calculate the vacuum energies for both the broken phase ($\sigma\neq 0$)
   and the symmetric phase ($\sigma=0$).
Unfortunately, the result becomes zero in both cases, and we cannot judge
   which vacuum should be selected.
One of the reasons for this will be the peculiar situation in the Hamiltonian
   of the Gross-Neveu model.
Since the classical Hamiltonian becomes zero in the Gross-Neveu model,
   the vacuum energy is also zero.
This peculiarity might affect the above result even in the quantum level.
Furthermore, there is the possibility that a different operator ordering 
   will resolve this problem.
Here, we do not go into this problem any further 
   but treat only the broken phase.

 From a viewpoint of the constituent quark picture, it is important to divide
   the light-cone Hamiltonian into a nonperturbative part and a
   perturbative part so that the nonperturbative part gives the kinetic term
   of the fermion with the ``constituent mass" \cite{rf:WilRob}. \ 
This is simply realized by use of the nontrivial solution  Eq.\ (\ref{cu})
   of the zero-mode constraint.
$P^-$ (up to leading order in $1/N $ ) is then represented as
\begin{equation}
  P^- = P_{\rm nonpert}^- + P_{\rm pert}^- \; ,
  \label{dj}
\end{equation}
where
\begin{eqnarray}
  P_{\rm nonpert}^- & = & \sum_{a=1}^{N} ~ \sum_{n= \frac{1}{2} }^{\infty}
          \left( \frac{\sigma^2}{k_n^+} \right)
        ( ~ b_n^{\, a \dag} b_n^{\,a} + d_n^{\, a \dag} d_n^{\,a}  ~) ~,
             \nonumber   \\
  P_{\rm pert}^- & = & \frac{L}{4 \pi}~
        \sum_{a=1}^{N} ~ \sum_{n= \frac{1}{2} }^{\infty}
        \left\{ ( \phi_0 )_{\rm oper}^{\: 2} + 2 ~\sigma ~
        ( \phi_0 )_{\rm oper} \right\}        \nonumber  \\
   & & \hskip 2.3cm    \times ( ~\frac{1}{n}~ b_n^{\, a \dag} b_n^{\,a}
        +~\frac{1}{n}~ d_n^{\, a \dag} d_n^{\,a}  ~) + \rm{h.c.}
  \label{dk}
\end{eqnarray}
The first term $  P_{\rm nonpert}^- $ represents the free fermions with
   dynamically generated mass $\sigma$, Eq.\ (\ref{cu}) ( ``constituent mass" ),
   and the second term $ P_{\rm pert}^- $
   denotes an interaction term, in which the operator part
   $ ( \phi_0 )_{\rm oper} $ is given by solving the zero-mode constraint
   Eq.\ (\ref{cd}) .
The excitation state $ b_n^{\, a \dag} | 0 \rangle $ describes the fermion
   with mass $\sigma$ and momentum $ k_n^+ = 2 \pi n /L $.
Note that we cannot obtain the nonperturbative part $  P_{\rm nonpert}^- $ 
   explicitly without solving the zero-mode constraint to
   leading order in $1/N$.
%
%
%
%
%
%
\section{Conclusion}

The (generalized) Gross-Neveu model was studied in order to see
   how spontaneous symmetry breaking is realized in the framework
   of the discretized light-cone quantization, and to examine the constituent
   quark picture of the model by use of the light-cone Hamiltonian.

The light-cone theory is a constraint system, so there arises the
   constraint equation involving the zero mode, according to the Dirac
   method.
That constraint equation, called the zero-mode constraint, determines
   the zero mode in terms of other oscillation modes.
We solved the zero-mode constraint to leading order in $1/N$.
In the constraint, the summation of the light-cone momentum
   $ 1/k_n^+ $ has an infrared divergence as well 
   as the ultraviolet divergence.
In order to avoid this divergence, we have introduced the damping factor 
   $ f( k^+ ) $ which involved the ultraviolet cutoff $ \Lambda'$ and the
   observed fermion mass $\sigma$.
The proper treatment of the infrared divergence is important
   to obtain a nonzero vacuum expectation value $\sigma$
   of the zero mode representing the spontaneous discrete chiral symmetry
   breaking.

We have shown that the vacuum state $|0 \rangle $ is an eigenstate of both
   the light-cone momentum $P^+ $ and the light-cone Hamiltonian $P^-$
   even in the case of spontaneous symmetry breaking
   (the problem of operator ordering remains).
These properties are important when one constructs the constituent quark
   picture.
Solving the zero-mode constraint to leading order
   in $1/N$, we can devide the light-cone Hamiltonian into two parts.
One is the
   kinetic term of the fermion with ``constituent mass" $\sigma $,
   and the other is the interaction term of fermions, whose explicit form
   will be given after we solve the zero-mode constraint up to next 
   leading order.

Finally, we hope that the study of the light-front Gross-Neveu model can
   be helpful to understand the light-front QCD.
\vskip 1.5 cm
\centerline{\bf Acknowledgements}\par
One of the authors (S.M.) would like to acknowledge 
   the kind hospitality of the High Energy
   Theory Group at Tokyo Metropolitan University.
He is also grateful to
   the members of the High Energy Theory Group and Doyo-kai
   for stimulating
discussions. 
The other author (K.I.) acknowledges Dr. T. Heinzl, Dr. K. Harada,
   and Dr. M. Taniguchi for various inspiring discussions.
%
%
%
%
%
%

%
%

\vskip 1cm
\begin{thebibliography}{99}
%
%
\bibitem{rf:Dir}
   P.~A.~M.~Dirac, Rev.~Mod.~Phys.\ {\bf 21} (1949), 392.
\bibitem{rf:rev}
   For a review, see
   S.~J.~Brodsky and H.-C.~Pauli,
      in {\it Recent Aspects of Quantum Fields},
      ed. H.~Mitter and H.~Gausterer, {\it Lecture Notes in Physics} 
      (Springer-Verlag, Berlin, 1991), vol. 396. \ 
   S.~J.~Brodsky, G.~McCartor, H.-C.~Pauli and S.~S.~Pinsky,
      Part.~World.\ {\bf 3} (1993), 109.\  
   K.~G.~Wilson, T.~S.~Walhout, A.~Harindranath, W.-M.~Zhang, R.~J.~Perry
      and S.~D.~Glazek, Phys.~Rev.\ {\bf  D49} (1994), 6720.
\bibitem{rf:HorBroPau}
   K.~Hornbostel, S.~J.~Brodsky and H.-C.~Pauli,
      Phys.~Rev.\ {\bf D41} (1990), 3814.
\bibitem{rf:WilRob}
    K.~G.~Wilson and D.~G.~Robertson,
       {\it Theory of Hadrons and Light-Front QCD}, ed. St.~D.~Glazek
       (World Scientific, Singapore, 1995), p.15.
\bibitem{rf:HeiKruSimWer}
    T.~Heinzl, S.~Krusche, S.~Simb\"urger and E.~Werner,
      Z.~Phys.\ {\bf  C56} (1992), 415.
\bibitem{rf:Rob}
    D.~G.~Robertson, Phys.~Rev.\ {\bf D47} (1993), 2549.
\bibitem{rf:PinSanHil}
    S.~S.~Pinsky, B.~van~de~Sande and J.~R.~Hiller,
     Phys.~Rev.\ {\bf  D51} (1995), 726.
\bibitem{rf:BorGraWer}
    A.~Borderies, P.~Grang\'e and E.~Werner,
    Phys.~Lett.\ {\bf B345} (1995), 458.
\bibitem{rf:Hor}
    K.~Hornbostel, Phys.~Rev.\ {\bf  D45} (1992), 3781.
\bibitem{rf:ThiOht}
    M.~Thies and K.~Ohta, Phys.~Rev.\ {\bf  D48} (1993), 5883.
\bibitem{rf:Pes}
   I.~Pesando, Mod.~Phys.~Lett.\ {\bf  A10} (1995), 525, 2339.
\bibitem{rf:KojSakSak}
   S.~Kojima, N.~Sakai and T.~Sakai,
        Prog.~Theor.~Phys.\ {\bf 95} (1996), 621.
\bibitem{rf:Mae}
   S.~Maedan, Phys.~Lett.\ {\bf  B370} (1996), 116.
\bibitem{rf:PauBro}
   H.-C.~Pauli and S.~J.~Brodsky,
    Phys.~Rev.\ {\bf  D32} (1985), 1993, 2001.
\bibitem{rf:MasYam}
   T.~Maskawa and K.~Yamawaki,
          Prog.~Theor.~Phys.\ {\bf 56} (1976), 270.
\bibitem{rf:KalPauPin}
   A.~C.~Kalloniatis, H.-C.~Pauli and S.~Pinsky,
                    Phys.~Rev.\ {\bf  D50} (1994), 6633.
\bibitem{rf:GroNev}
   D.~J.~Gross and A.~Neveu, Phys.~Rev.\ {\bf  D10} (1974), 3235.
\bibitem{rf:Kug}
   T.~Kugo, Soryushiron Kenkyu (Kyoto) {\bf 53} (1976), 1.
\bibitem{rf:DirLec}
   P.~A.~M.~Dirac, {\it Lectures on Quantum Mechanics},
                 Belfer Graduate School of Science, 
                 Monograph Series ( New York, 1964 ).
\bibitem{rf:ItaMae}
   K.~Itakura and S.~Maedan, in preparation.
\bibitem{rf:DieHeiSchWer}
   C.~Dietmaier, T.~Heinzl, M.~Schaden and E.~Werner,
   Z.~Phys.\ {\bf A334} (1989), 215.
\bibitem{rf:HarVar}
   A.~Harindranath and J.~P.~Vary, Phys.~Rev.\ {\bf D37} (1988), 3010.
\bibitem{rf:NakYam}
   N.~Nakanishi and K.~Yamawaki, Nucl.~Phys.\ {\bf B122} (1977), 15.
%
\end{thebibliography}
\end{document}